\documentclass[a4paper,fleqn,usenatbib]{mnras}

\usepackage{newtxtext,newtxmath}

\usepackage[T1]{fontenc}
\usepackage{ae,aecompl}

\usepackage{graphicx}	
\usepackage{amsmath}	
\usepackage{amssymb}	

\title[Gamma Rays from Supernova Remnants]{Time Evolution of Gamma Rays from Supernova Remnants}

\author[D. Gaggero et al.]{
Daniele Gaggero,$^{1}$\thanks{E-mail: D.Gaggero@uva.nl}
Fabio Zandanel,$^{1}$
Pierre Cristofari,$^{2}$
and Stefano Gabici$^{3}$
\\
$^{1}$GRAPPA, Institute of Physics, University of Amsterdam, 1098 XH Amsterdam, The Netherlands\\
$^{2}$Columbia University, 10027 New York, USA\\
$^{3}$APC, Univ. Paris Diderot, CNRS/IN2P3, CEA/Irfu, Obs. de Paris, Sorbonne Paris Cit\'e, 75013 Paris, France
}

\pubyear{2017}

\begin{document}
\label{firstpage}
\pagerange{\pageref{firstpage}--\pageref{lastpage}}
\maketitle

\begin{abstract}

{We present a systematic phenomenological study focused on the time evolution of the non-thermal radiation -- from radio waves to gamma rays -- emitted by typical supernova remnants via hadronic and leptonic mechanisms, for two classes of progenitors: thermonuclear and core--collapse. 

To this aim, we develop a numerical tool designed to model the evolution of the cosmic-ray spectrum inside a supernova remnant, and compute the associated multi--wavelength emission. We demonstrate the potential of this tool in the context of future population studies based on large collection of high-energy gamma-ray data.

We discuss and explore the relevant parameter space involved in the problem, and  focus in particular on their impact on the maximum energy of accelerated particles, in order to study the effectiveness and duration of the PeVatron phase. 

We outline the crucial role of the ambient medium through which the shock propagates during the remnant evolution. In particular, we point out the role of dense clumps in creating a significant hardening in the hadronic gamma-ray spectrum.}

\end{abstract}




\section{Introduction}

Understanding the nature of Galactic sources capable of accelerating hadronic cosmic rays (CRs) all the way up to $\mathcal{O}(PV)$ rigidities is one of the main goals in high energy atrophysics.
The crucial role of supernova remnants (SNRs) in this context was first proposed in \cite{Baade1934} out of general energy budget arguments; two decades later the picture was better defined in term of SNRs located {\it in our Galaxy} \citep{TerHaar1950, morrison1957, ginzburg1956} -- although a mechanism able to provide an effective CR acceleration at SNR shocks had not been proposed yet at that time. 
Other classes of sources may also be at work in the Galactic environment, for example OB associations may significantly contribute, as first noticed in the 1970s \citep{reeves1973,montmerle1979}, and recently pointed out again by \cite{murphy1916} in light of the elemental compositions measured by balloon experiments such as super-TIGER. However, the SNR paradigm, though not proven, remains the most widely studied and discussed in the literature.

A detailed theory of SNR diffusive shock acceleration was developed by Bobalsky\footnote{Thierry Montmerle's abbreviation for Blandford, Ostriker, Bell, Axford, Leer, Skadron and Krymsky.}  \citep{ostriker1978,bell1978,icrc1977,krymskii1977}, and it was soon realized that a significant amplification of the magnetic field at the shock is required to reach PeV energies. The origin of this amplification is still unknown and is currently the major problem to be solved in this theoretical framework -- see, in particular, \cite{bell04}, \cite{drury2012}, \cite{giacalone2007}, and the recent discussion in \cite{gabici2016}.
In order to test the theory and identify whether SNRs are actually able to accelerate hadrons up to PV rigidities, and how long this process can be sustained, it is useful to investigate the broad-band non-thermal radiation emitted by these objects, from radio waves all the way up to gamma rays.


On the observational side, we are gathering many interesting data at all wavelengths.
The current generation of Cherenkov gamma-ray telescopes (H.E.S.S., MAGIC and VERITAS) have detected a population of $16$ TeV sources associated with shell-type SNRs\footnote{\url{http://tevcat.uchicago.edu}}, while few tens of firm identifications at GeV energies are listed in the first Fermi catalogue of SNRs \citep{2016ApJS..224....8A}. However, most of these objects are quite old and none of these -- including 
the very bright, well-studied RX J$1713.7-3946$ (see, e.g., \citealt{gabici2014} and references therein) -- show striking evidence of ongoing PeV acceleration. 
The only claim of a gamma-ray emission with no clear cutoff in the multi-TeV domain comes from the Galactic center region \citep{pevatron2016}. That discovery triggered a debate in the community \citep{gaggero2017} \citep{terrier2017} about the origin and the nature of the observed emission.

Among the youngest SNRs observed, the most studied are SN1572 (Tycho's nova, see, e.g., the TeV data from \citealp{acciari2011} and the analysis from \citealp{morlino2012}), SN1604 (Kepler's nova, see \citealp{vink2016} for a review), and Cassiopea A \citep{acciari2010,zirakashvili2014}. All those sources are more than 3 centuries old and do not show clear evidence for ongoing PeV acceleration of protons. The much younger G1.9$+$0.3 \citep{reynolds2008} is $\sim 8$ kpc away and embedded in the bright Galactic ridge region, so it is difficult to identify its high-energy gamma-ray emission. The youngest known remnant in the local group of galaxies, SN1987, is a type II remnant located even farther away, in the Large Magellanic Cloud \citep{SN1987}, and therefore only future, more sensitive, Cherenkov gamma-ray telescopes such as the Cherenkov Telescope Array (CTA; \citealp{CTA}) will have the opportunity to study its spectrum up to multi-TeV energies.

On the modeling side, despite this large collection of data, a clear comprehension of the crucial aspects of CR acceleration outlined above is still lacking. 
The goal of this paper is to provide a systematic study about the evolution of the gamma-ray emission from SNRs as a function of time, with the future prospect of a detailed comparison with an even more extended and accurate set of data regarding a large population of young and middle-aged SNRs.
To this aim, we develop a numerical tool designed to compute the maximum energy of accelerated particles as a function of a well-defined set of free parameters, follow the evolution of CR spectra accordingly, and finally provide a prediction for the radio to gamma-ray spectra at different times.
{This tool includes a detailed description of the ambient medium where the shock propagates. We pay particular attention to the duration of the PeVatron phase and outline, in this context, the crucial role of the high-density progenitor star wind. Moreover, we discuss how dense clumps may significantly shape the spectrum of the hadronic emission. }


The paper is organized as follows. In Section 2 we describe our setup for the time evolution of a type I and II SNRs. In Section 3 we discuss how the maximum energy of hadronic and leptonic acceleration evolves with time. Section 4 presents our framework for the time evolution of the CR spectrum in the remnant. In Section 5 we discuss the resultant broadband spectra, with particular emphasis on the gamma-ray emission and its time evolution as a function of the free parameters considered. Finally, in Section 6 we present our conclusions.

\section{Dynamical evolution of supernova remnants}
\label{sec:SNRs}

The dynamical evolution of a SNR shell can be divided into three temporal phases: the ejecta dominated, the Sedov, and the radiative phase. Here, we restrict our attention to the two earlier phases, where most of the acceleration of CRs is believed to happen. The transition between the ejecta dominated and Sedov phase happens when the mass of the interstellar gas swept up by the shock equals the mass of the stellar matter ejected in the supernova explosion.

For a type Ia supernova exploding in a homogeneous medium, the time evolution of the radius and velocity of the shock in the ejecta dominated phase can be written as \citep{chevalier1982}:
\begin{eqnarray}
\label{eq:Ia}
R_s &=& 5.3 \left( \frac{E_{51}^2}{M_{ej,\odot} n} \right)^{1/7} t_{\rm kyr}^{4/7} ~ {\rm pc} \, , \\ 
u_s &=& 3.0 \times 10^3 \left( \frac{E_{51}^2}{M_{ej,\odot} n} \right)^{1/7} t_{\rm kyr}^{-3/7} ~ {\rm km/s} \, , 
\end{eqnarray}
where $E_{51}$ is the supernova explosion energy in units of $10^{51}$ erg, $M_{ej, \odot}$ the mass of the ejecta in solar masses, and $n$ the density of the interstellar medium in cm$^{-3}$. In deriving the expressions above it has been assumed that the ejecta have a radial power law density profile $\varrho \propto r^{-k}$ with $k = 7$ \citep{chevalier1982}.
On the other hand, the shock resulting from a type II supernova initially propagates in the wind of the progenitor star, characterised by a density profile $\varrho_w = \dot{M}/(4 \pi u_w r^2) $ where $\dot{M}$ and $u_w$ are the mass loss rate and the velocity of the wind, respectively.
The time evolution of the radius and velocity of the SNR shock can then be written as \citep{chevalierliang}:
\begin{eqnarray}
R_s &=& 7.7 \left( \frac{E_{51}^{7/2} u_{w,6}}{\dot{M}_{-5} M_{ej,\odot}^{5/2}} \right)^{1/8} t_{\rm kyr}^{7/8} ~ {\rm pc}  \, ,  \\
u_s &=& 6.6 \times 10^3 \left( \frac{E_{51}^{7/2} u_{w,6}}{\dot{M}_{-5} M_{ej,\odot}^{5/2}} \right)^{1/8} t_{\rm kyr}^{-1/8} ~ {\rm km/s}  \, , 
\label{eq:II}
\end{eqnarray}
where $\dot{M} = 10^{-5} \dot{M}_{-5} ~ M_{\odot}$/yr and $u_w = 10^6 u_{w,6}$~cm/s and a density profile of the ejecta with $k = 10$ has been assumed.
For completeness, we mention that a simplified description of the ejecta dominated phase has been often adopted in the literature \citep[see, e.g.,][]{finke}. According to this picture the shock speed is assumed {to be constant during most of the ejecta-dominated phase}, and determined by the equation:
\begin{equation}
\label{eq:free0}
E_{SN} \sim \frac{1}{2} M_{ej} u_s^2  \, , 
\end{equation}
where $E_{SN}$ is the supernova explosion energy and $M_{ej}$ the mass of ejecta. The equation describes the fact that in the very early phase of the evolution of a SNR, the energy is virtually all in the form of the kinetic energy of the ejecta. This gives a constant value of the shock speed equal to:
\begin{equation}
\label{eq:free}
u_s = 10^4 E_{51}^{1/2} M_{ej,\odot}^{-1/2} ~ {\rm km/s}  \, .
\end{equation}
{In the following we will consider both the self similar scalings given by Equations \ref{eq:Ia}-\ref{eq:II} and the simplified expressions given by Equation \ref{eq:free} for estimating the maximum energies attained by electrons and protons (see Fig.~\ref{fig:EmaxIa}). However, we adopt only the self similar scalings of Equations \ref{eq:Ia}-\ref{eq:II} in the further calculation of the evolution of CR spectra and of the corresponding non-thermal radiations.}

To model the evolution of the shock radius and velocity in the Sedov phase we adopt the thin-shell approximation, which has been widely used to describe the dynamical evolution of SNRs \citep[see, e.g.,][]{ostrikermckee,bisnovatyi}.
It is based on the assumption that the mass in the SNR is concentrated within a spherical shell of zero thickness located at the forward shock, at a radial coordinate $r = R_s$.
The gas in the shell moves at the velocity of the downstream fluid, which for a strong shock is $(3/4)~u_s$.
For non radiative systems, the total energy is conserved and is equal to the supernova explosion energy $E_{\rm SN}$. It is the sum of the kinetic energy of the shell and the SNR thermal energy:
\begin{equation}
\label{eq:energy}
E_{\rm SN} = \frac{1}{2} M u_s^2 + \frac{4 \pi}{3} R_s^3 \frac{P_{in}}{\gamma+1}  \, , 
\end{equation}
where $P_{in}$ is the gas pressure in the SNR, $\gamma$ the gas adiabatic index, and $M$ the mass of the shell:
\begin{equation}
\label{eq:mass}
M = M_{ej} + 4 \pi \int_0^{R_s} \varrho(r) r^2 {\rm d}r  \, , 
\end{equation} 
where the first term on the right-hand side represents the mass of the supernova ejecta and the second the mass of the ambient medium (of mass density $\varrho$) swept up by the shock.
The equation of momentum conservation reads:
\begin{equation}
\label{eq:momentum}
\frac{\rm d}{{\rm d}t} \left( M u_s \right) = 4 \pi R_s^2 P_{in}  \, , 
\end{equation}
where we assumed that the SNR shock expands in a cold ambient medium of negligible pressure $P_0 \ll P_{in}$.
The three equations above define the dynamics of the SNR until the shock becomes radiative, and cooling cannot be neglected anymore \citep[e.g.,][]{cioffi}, or subsonic, and dissolves in the interstellar medium.

In order to solve Equations \ref{eq:energy}, \ref{eq:mass}, and \ref{eq:momentum} and obtain the time evolution of the SNR shock position and velocity, we need to specify the supernova explosion energy $E_{\rm SN} = 10^{51} E_{51}$~erg, the mass of the ejecta $M_{ej}$, and the profile of the ambient density around the supernova, $\varrho(r)$.
As discussed above, for a thermonuclear (type Ia) supernova we assume a homogeneous ambient medium of density $\varrho = \mu~ n ~m_p$ where $m_p$ is the proton mass and $\mu m_p \sim 1.4 m_p$ the mean interstellar atom mass per H nucleus. 
On the other hand, for a core-collapse (type II) supernova the SNR shock propagates first through the red supergiant wind $\varrho \propto r^{-2}$. After that, in many cases the propagation proceeds through a tenuous bubble inflated by the wind of the progenitor star in main sequence, especially for the most massive progenitors, and finally in the interstellar medium \citep[see, e.g.,][]{ptuskinzirakashvili2005,pierre,pierre2,dwarkadas}.
The typical radius of the wind  $R_{\rm{w}}$ is estimated by equating the ram pressure of the wind itself $P_{\rm ram}= \dot{M} u_{\rm w}/ (4 \pi r^2)$ to the thermal pressure of the bubble interior \citep[see, e.g.,][]{etienne} and is of the order of few parsecs.
Finally, the radius of the tenuous hot bubble is $R_{\rm b} = 28 (L_{36}/n_0)^{1/5} t_{\rm Myr}^{3/5}$~pc, where $L_{36}$ is the main--sequence star wind power in units of $10^{36}$~erg/s, $n_0$ the density of the interstellar medium outside of the bubble, and $t_{\rm Myr}$ is the wind lifetime in units of mega--years. The density inside the bubble is $n_{\rm b} = 0.01 (L_{36}^{6} n_0^{19} t_{\rm Myr}^{-22})^{1/35}$~cm$^{-3}$ and the gas temperature $T_b = 1.6 \times 10^6 (L_{36}^{8} n_0^{2} t_{\rm Myr}^{-6})^{1/35}$~K \citep{mccray,weaver}. Here we assume the wind lifetime to be of the order of several Myr, which corresponds to the duration of the main sequence phase of very massive stars \citep{longair}. 

\section{Maximum energy of accelerated particles}

As mentioned in the Introduction, it is a well known fact that the acceleration of CRs up to PeV energies at SNR shocks requires a significant amplification of the magnetic field \citep[e.g.,][]{lagagecesarsky,hillas}.
Indeed, this picture is supported by X-ray observations of several young SNRs, which revealed magnetic fields whose strengths are well in excess of the typical interstellar value of a few microGauss  \citep[e.g.,][]{jacco,yas}.
\citet{bell04} suggested that the field amplification might be due to a plasma instability induced by the streaming of CR protons away from shocks. 
Another possibility is to amplify the field as the result of the turbulence induced
by the CR gradient upstream of the shock acting on an inhomogeneous ambient medium. This mechanism is often referred to as Drury instability (see, e.g., \citealt{drury2012}, and references
therein).

The effectiveness of such mechanisms in amplifying the magnetic field to the level needed to accelerate CRs up to the PeV domain is still debated (see, e.g., the reference list in \citealp{gabici2016}), but observations of young SNRs suggest that few percent of the shock ram pressure is indeed converted into magnetic field pressure downstream of the shock \citep{heinz}. In the following, following \citet{heinz}, we assume $\xi_B \sim 3.5$\% as a reference value. The reader should keep in mind that this is a very uncertain value, 
and that smaller (by a factor of a few) values of $\xi_B$ have been estimated based on theoretical studies \citep[e.g.][]{bell2013}. As we will show in the following, this parameter is crucial in determining the maximum energy of accelerated particles, which scales as $\propto \xi_B^{1/2}$. 

We now develop a recipe to estimate the maximum energy of particles acelerated at SNR shocks as a function of the SNR age.
Particles are accelerated at SNRs as the result of repeated cycles around the shock. After each cycle, a test particle gains a momentum \citep{luke}:
\begin{equation}
\label{eq:cycle}
\Delta p = \frac{4}{3} \frac{u_1-u_2}{v} p \, , 
\end{equation} 
where $p$ is the particle momentum, $v$ its speed, and $u_{1(2)}$ is the fluid velocity upstream (downstream) of the shock measured in the shock rest frame.
The time needed to complete a cycle is:
\begin{equation}
\Delta t = \frac{4}{v} \left( \frac{D_1}{u_1}+\frac{D_2}{u_2} \right)  \, , 
\end{equation}
expressed as the sum of the up- and downstream residence time of a particle. 
Here, $D_{1(2)}$ is the particle diffusion coefficient, assumed to be spatially uniform upstream (downstream) of the shock.
Shocks are very turbulent environments and thus particle diffusion is likely to proceed at the Bohm rate, $D = (1/3) r_L v$, where $r_L = pc/qB$ is the Larmor radius of a particle in a magnetic field of strength $B$, and $q$ is the elementary charge. At a strong shock of compression factor $r = u_1/u_2 = 4$, a turbulent magnetic field is compressed, on average, by a factor of $\sigma = \sqrt{11}$, implying an instantaneous acceleration rate of:
\begin{equation}
\label{eq:accrate}
\frac{{\rm d}p}{{\rm d}t} = \frac{\Delta p}{\Delta t} \sim 0.11 \frac{u_s^2}{D_1} p  \, .
\end{equation}
Strictly speaking, Equation \ref{eq:accrate} refers to a shock of constant velocity, but following \citet{lagagecesarsky} we will use it also in the more general case of a shock with a velocity dependent on time.

The maximum energy of particles accelerated at the shock of a SNR of age $t_{age}$ is given by the most stringent among the following conditions.

\begin{itemize}
\item{{\it Age limited acceleration}. Equation~\ref{eq:accrate} can be integrated up to a time $t_{age}$ and give the maximum energy of accelerated particles in the absence of particle energy losses and escape from the system \citep{lagagecesarsky}.}\\
\item{{\it Escape limited acceleration}. The maximum energy due to particle escape is estimated by equating the diffusion length of particles ahead of the shock $l_d = u_s/D$, to some fraction, typically $\chi \approx 0.05 ... 0.1$ of the SNR shock radius \citep[see, e.g.,][]{ptuskinzirakashvili2005}. In the following we will adopt the value $\chi = 0.05$.}\\
\item{{\it Loss limited acceleration}. While the acceleration of protons at SNR shocks proceeds unimpeded by energy losses, the acceleration of electrons can be limited by synchrotron radiative losses acting at a rate ${\rm d}p/{\rm d}t |_s$. The maximum energy of accelerated electrons can then be derived by equating the momentum gain per cycle (Equation \ref{eq:cycle}) to the momentum loss per cycle, given by $\Delta p |_s = {\rm d}p/{\rm d}t |_{s,1} \Delta t_1 + {\rm d}p/{\rm d}t |_{s,2} \Delta t_2$, where the subscripts refer to the upstream (1) and downstream (2) region of the shock and $\Delta t_i = 4 D_i/v u_i$ with $i = 1,2$ \citep[see, e.g.,][]{giulia}.
}
\end{itemize}

We stress that the value of the magnetic field at the shock is a crucial parameter in order to determine the time evolution of the maximum energy of protons and electrons accelerated at SNRs.
Both observations \citep[see, e.g.,][]{heinz} and theory \citep[see, e.g.,][]{bell2013} suggest that the acceleration of particles at shocks induces an amplification of the field such that a small fraction $\xi_B$, at the few percent level, of the shock ram pressure is converted into magnetic pressure downstream of the shock: $\xi_B \varrho u_s^2 = B_2^2/ 8 \pi$. 
This implies that the upstream field is amplified up to a value of:
\begin{equation}
B_1 \approx 43 \left( \frac{\xi_B}{0.035} \right)^{1/2} \left( \frac{n}{\rm cm^{-3}} \right)^{1/2} \left( \frac{u_s}{1000~{\rm km/s}} \right) ~ \mu \rm G  \, , 
\end{equation}
when the shock speed is larger than:
\begin{equation}
u_* = \frac{\sigma B_0}{\sqrt{8 \pi \xi_B \varrho}} \approx 0.2 \times 10^8 \left( \frac{n}{\rm cm^{-3}} \right)^{-1/2} ~ \rm km/s  \, , 
\end{equation}
where $B_0 \approx 5 \mu \rm G$ is the value of the interstellar magnetic field and we have adopted $\xi_B \approx 0.035$ which has been inferred from observations \citep{heinz}.
For smaller velocities of the shock the magnetic field amplification is ineffective and $B_1 = B_0$.
To describe the diffusion coefficient of particles at this late phase ($u_s < u_*$) we follow the phenomenological approach by \citet{zirakashviliptuskin2012} and we multiply the expression of the Bohm diffusion coefficient by the factor $(1+(u_*/u_s)^2)^3$, which implies that the diffusion coefficient becomes larger (i.e., particles are less confined) as the shock slows down.

Within this framework, we can compute the maximum energy of particles accelerated at a given SNR at a given time. For protons, the maximum energy is equal to $\min(E_{max}^{age},E_{max}^{esc})$, while for electrons is equal to $\min(E_{max}^{age},E_{max}^{esc},E_{max}^{syn})$.

\begin{table}
	\centering
	\caption{Parameters adopted to describe the evolution of supernovae of type Ia and II. The rows refer to: the supernova explosion energy; the ejecta mass; the magnetic field amplification efficiency; the  ISM number density (relevant for SNIa); the red supergiant wind mass loss rate and velocity, the ISM number density in the cavity (relevant for SNII); the ratio between magnetic and total energy inside the remnant $\xi_{\rm inside}$. All quantities are expressed in normalised units as in Section~\ref{sec:SNRs}.}
	\label{tab:example_table}
	\begin{tabular}{cccccccc} 
		\hline
		parameter &  type Ia & type II \\
		\hline
		$E_{51}$ & 1 & 1 \\
		$M_{ej,\odot}$ & 1.4 & 3 \\
        $\xi_B$ & $3.5$\% & $3.5$\% \\
        $n$ & 0.1 & $-$ \\
        $\dot{M}_{-5}$ & $-$ & 2 \\
        $u_{w,6}$ & $-$ & 1 \\
        $n_{\rm cavity}$ & $-$ & 0.01 \\
        $\xi_{\rm inside}$ & 0.001 & 0.001 \\
		\hline
	\end{tabular}
\end{table}

\begin{figure*}
	\includegraphics[width=\columnwidth]{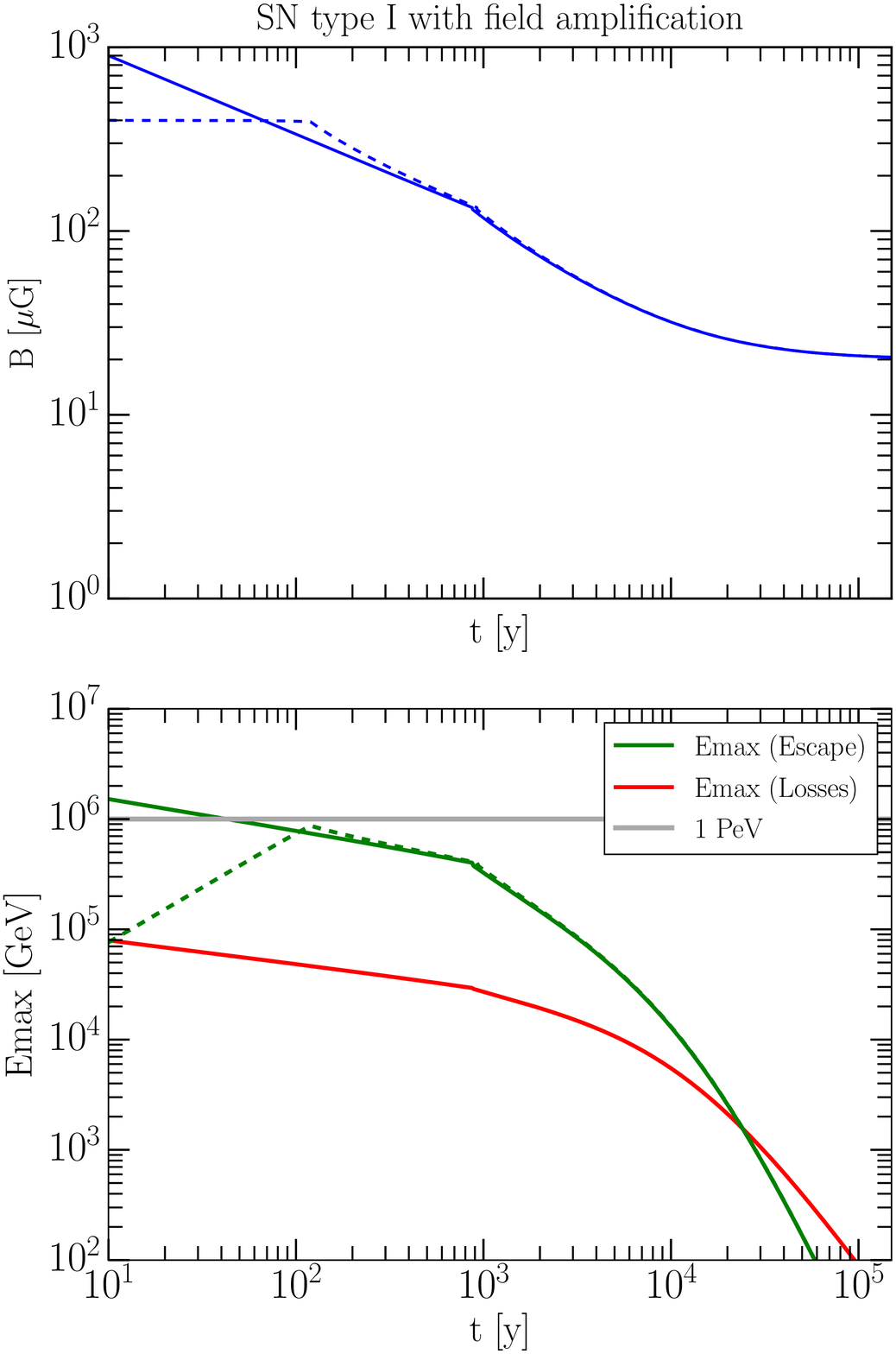}
	\includegraphics[width=\columnwidth]{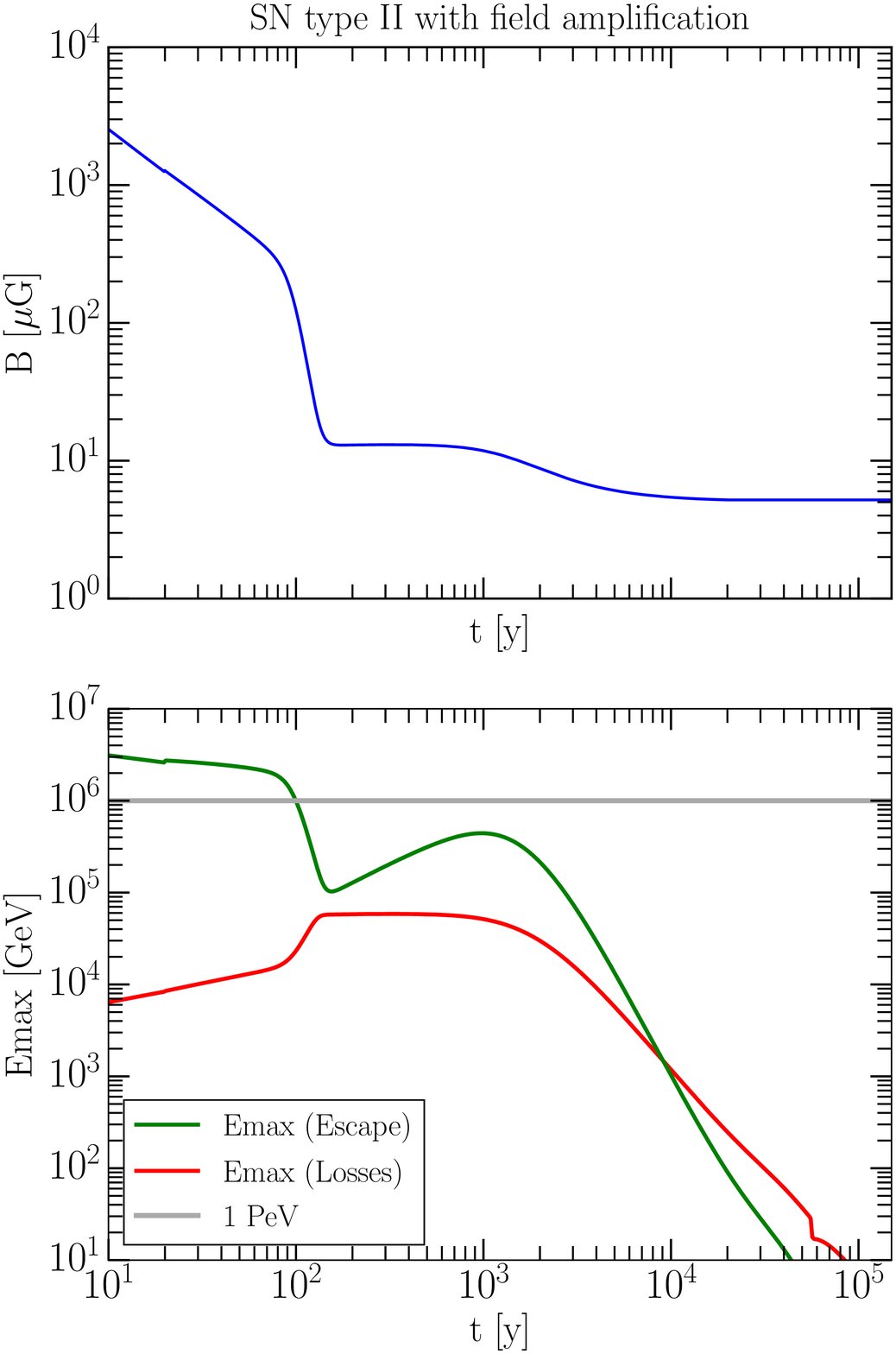}    
    \caption{{\bf Upper panel:} Magnetic field strength upstream of the shock  (see Table 1) as a function of the SNR age. {\bf Lower panel:} Maximum energy of accelerated particles at the shock. For CR protons the maximum energy is determined by particle escape from the shock (green lines), while for electrons the maximum energy is determined either by escape or by synchrotron losses (red lines). {In the type Ia case, solid (dashed) lines refer to the choice of Equations \ref{eq:Ia}-\ref{eq:II} (\ref{eq:free0}-\ref{eq:free}) to describe the ejecta-dominated phase. For the type II case, the scenario corresponding to eq. \ref{eq:free0}-\ref{eq:free} does not show any difference with respect to the other one for the range of times we are considering here.}}
    \label{fig:EmaxIa}
\end{figure*}


{The main result, for the default SNIa and SNII scenarios (see parameters in Table~\ref{tab:example_table}), is shown in Fig.~\ref{fig:EmaxIa}. 

The lower panels show the evolution of the maximum energy with time, while he upper panels represent the value of the magnetic field upstream of the SNR shock as a function of the SNR age. We remark that the age constraint is never relevant in the time span we are considering, therefore the corresponding line is not shown in the plot. The maximum energy for the protons, which do not suffer from IC and synchrotron energy losses, is thus simply given by $E_{max}^{esc}$ (green line), while -- for the leptonic component -- the most stringent constraint between $E_{max}^{esc}$ and $E_{max}^{loss}$ applies. 

The plots clearly outline, in agreement with previous findings \cite{bell2013}, the capability of a typical type II remnant to sustain a PeVatron phase during the first few decades of its evolution, when the shock is still very fast and is propagating in the thickest part of the wind, thus triggering a highly effective magnetic field amplification. 
 
Regarding type II remnants, we also remark how the structure of the ambient medium described above shapes the time evolution of both $E_{max}^{esc}$ and $E_{max}^{loss}$. In particular, the transition between the thick supergiant wind and the cavity is clearly visible at $\simeq 1$ century, and the second transition between cavity and ISM can be identified by the feature at $\simeq 8 \cdot 10^4$ y in the $E_{max}^{loss}$ evolution (red line).
 }

\section{Evolution of the cosmic-ray spectrum}

Given the scenario discussed above, we model the evolution of both the leptonic and hadronic CR spectrum inside the remnant.
We follow the SNR evolution for $\sim 10^5$ years, with a timestep of $\simeq 1$ year.
At each timestep, we inject electrons and protons according to an unbroken power-law 
\begin{equation}
Q = Q_0 E^{-\alpha}  \, , 
\end{equation}
up to $E_{\rm cut}$, where we fix $\alpha = 2.3$.

In particular, we recall that, for protons, the escape-limited maximal energy is dominant; for the electrons, instead, the minimum value between the loss-dominated and escape-dominated $E_{\rm max}$ has to be taken into account.
The normalization term $Q_0$ is set at $10\%$ of the incoming energy flux $f$, defined as $f = 1/2 \rho u_s^3 \cdot 4\pi R_s^2$. 

We numerically solve the time evolution of the electron spectrum inside the SNR following \cite{finke}:
\begin{equation}
\frac{\partial N(E,t)}{\partial t} \,+\, \frac{\partial}{\partial E} { \dot{E} N(E,t)} \,=\, Q(E,t)  \, , 
\end{equation}
where:
\begin{itemize} 
\item  for the protons, $\dot{E}$ is simply the adiabatic loss term $\dot{E} = k_{\rm ad} E / t$, with $k_{\rm ad} = 1$; 
\item for the electrons, besides adiabatic losses, we consider the dominant loss terms due to synchrotron emission and inverse Compton scattering on the CMB photons \cite{finke}. 
\end{itemize}

We discretize the above equation on a two-dimensional grid, with linear spacing in time and log-spacing in energy. 
A second-order implicit scheme is implemented, in order to guarantee a stable solution regardless of the chosen timestep (see, e.g., the 
discussion in \citealp{evoli2016} for the accuracy of this kind of scheme applied to CR energy losses).

In order to mimic the loss of particles upstream when the escape condition $l_D \,>\, \chi R_{\rm shock}$ discussed above is fulfilled, we apply {\it a posteriori} to the solution:
\begin{itemize}
\item In the hadronic case, an exponential cutoff at $E_{\rm max}$(escape) for each timestep $i$.\footnote{For the protons, we expect the maximum energy to be given by $E_{\rm max}$(escape) for most of the evolution, and to be monotonically decreasing with time. Therefore, this prescription correctly accounts for the fact that PeV protons accelerated in the early stages are lost during the subsequent evolution, when the magnetic field is lower and the Larmor radius is larger.} 
\item In the leptonic case, a super-exponential cutoff at $E_{\rm max}$(loss), when synchrotron losses dominate, and an exponential cutoff at $E_{\rm max}$(escape) at late times, when escape is most relevant (see fig. \ref{fig:EmaxIa}).
\end{itemize}

As far as the electron evolution is concerned, the key parameter in our problem is the mean magnetic field inside the remnant $B_{\rm inside}$, which determines the synchrotron energy-loss rate. 
In order to estimate the mean field, we assume that the total magnetic energy is a small fraction ($0.1\%$) of the internal energy. The prescription we adopt to compute the internal energy is the following.

\begin{itemize}
\item During the early ejecta-dominated phase, we compute the internal energy by integrating the self-similar pressure profile computed in \cite{chevalier82} and assuming an equation of state $\epsilon \,=\, 3/2 P$.
\item After the transition to Sedov, we compute the internal energy -- assuming the thin-shell approximation -- as follows:
\begin{equation}
E_{int} = E_{\rm SN} - \frac{1}{2} M u_S^2 \, , 
\end{equation}
with the same notation as Equation~\ref{eq:energy}. We set $E_{\rm SN} = 10^{51} {\rm erg}$. $M$ is the total (swept + ejecta) mass.
\end{itemize}

In general, we consider the interpolation between the two extreme regimes. 

Given this recipe, we compute consistently the average magnetic field inside the remnant, and therefore synchrotron energy-loss term for the electrons at all times. The final result of our computation is the complete time evolution of the electron and proton spectra. We remark that a clear {\it cooling break} naturally appears in the leptonic spectra. The position of this spectral feature is consistent with the theoretical prediction, i.e., $E_{break}(t) \simeq k_{synch} t$.
 

\subsection{Evolution in a Clumpy Medium}
When considering the case of type II supernovae, we additionally model the possibility that the SNRs expand in a clumpy medium. In this case, the spectrum of protons locked in clumps is much harder than the one obtained at the shock with the consequence that hadronic emission can result in harder spectra, similar to what typically obtained leptonically. See \cite{gabici2014} and references therein.

The interaction of the SNR shock with clumps can be modeled as follows~\citep[see, e.g.,][for an extensive description of the procedure]{gabici2014}.
We consider that the SNR shock is evolving in a clumpy medium where the clump density is $n_{\rm clump}\approx 3 $~pc$^{-3}$, and the gas density inside each clump is typically $10^{3}$ cm$^{-3}$. 
Once a clump enters the SNR shock, it is bombarded by the CR contained in the SNR shell, and the evolution of the total CR content inside the clump $N_{\rm cl}$ is described by the equation~\ref{eq:clump}:
\begin{equation}
\frac{\partial N_{\rm cl} (E) }{\partial t} = \frac{(V_{\rm cl}/V_{\rm s}) N_{\rm CR} (E) - N_{\rm cl}(E)}{\tau_{\rm d}} \, ,
\label{eq:clump}
\end{equation}
where $N_{\rm CR}$ is the CR spectrum inside the SNR shell,  $V_{\rm cl}$ and $V_{\rm s}$ are respectively  the volumes of the clump and of the SNR shell filled with CRs, and $\tau_{\rm d}$ the time needed for a CR to diffusive into a clump.
$V_{\rm cl}$ is estimated by assuming a spherical shape of radius $L_{\rm c} = 0.1$~pc. $V_{\rm s}$ is calculated considering most of the CR content remain between the forward shock $R_{\rm s}$ and the contact discontinuity at $\approx 0.9 \; R_{\rm s}$. 
The penetration time $\tau_{\rm d}$ can be estimated by considering that the diffusion of CRs in the very turbulent structure of the surrounding of the clumps is of the Bohm type and writing $\tau_{\rm d} \approx L_{\rm tr}^{2}/6 D_{\rm B}$ where $L_{\rm tr} =0.05$~pc is the thickness of the layer surrounding the clump and $D_{\rm B}$ the Bohm diffusion coefficient.

\section{Gamma-ray spectra and evolution}

\begin{figure*}
\begin{center}
\minipage{0.48\textwidth}
 \includegraphics[scale=0.56]{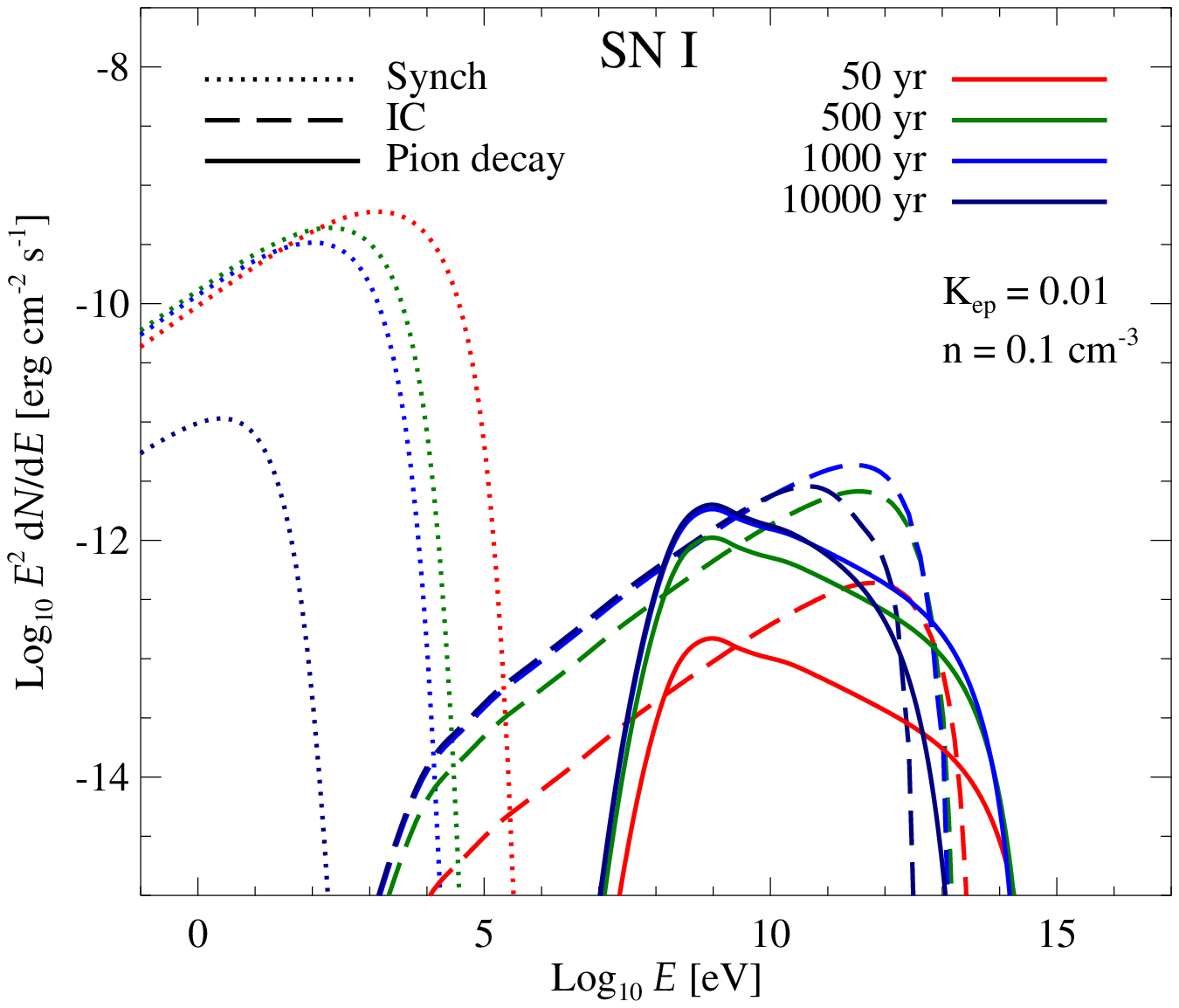}
\endminipage\hfill
\minipage{0.48\textwidth}
\includegraphics[scale=0.56]{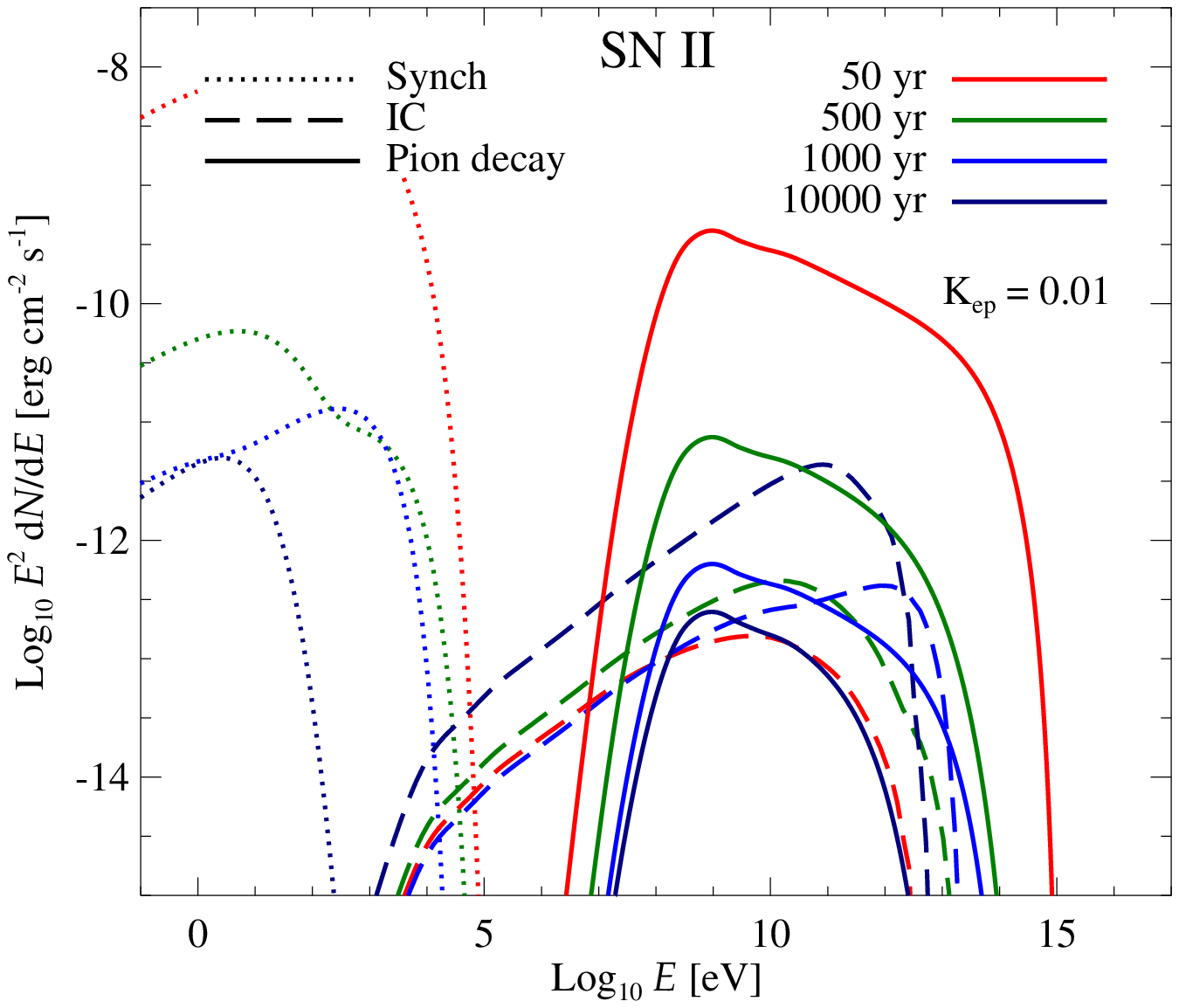}
\endminipage
\caption{Typical evolution of the broadband spectra of a type I and II SNRs for our reference values of the relevant parameters.}
\label{fig:benchmarkCases}
\end{center}
\end{figure*}

\begin{figure*}
\begin{center}
\minipage{0.48\textwidth}
 \includegraphics[scale=0.56]{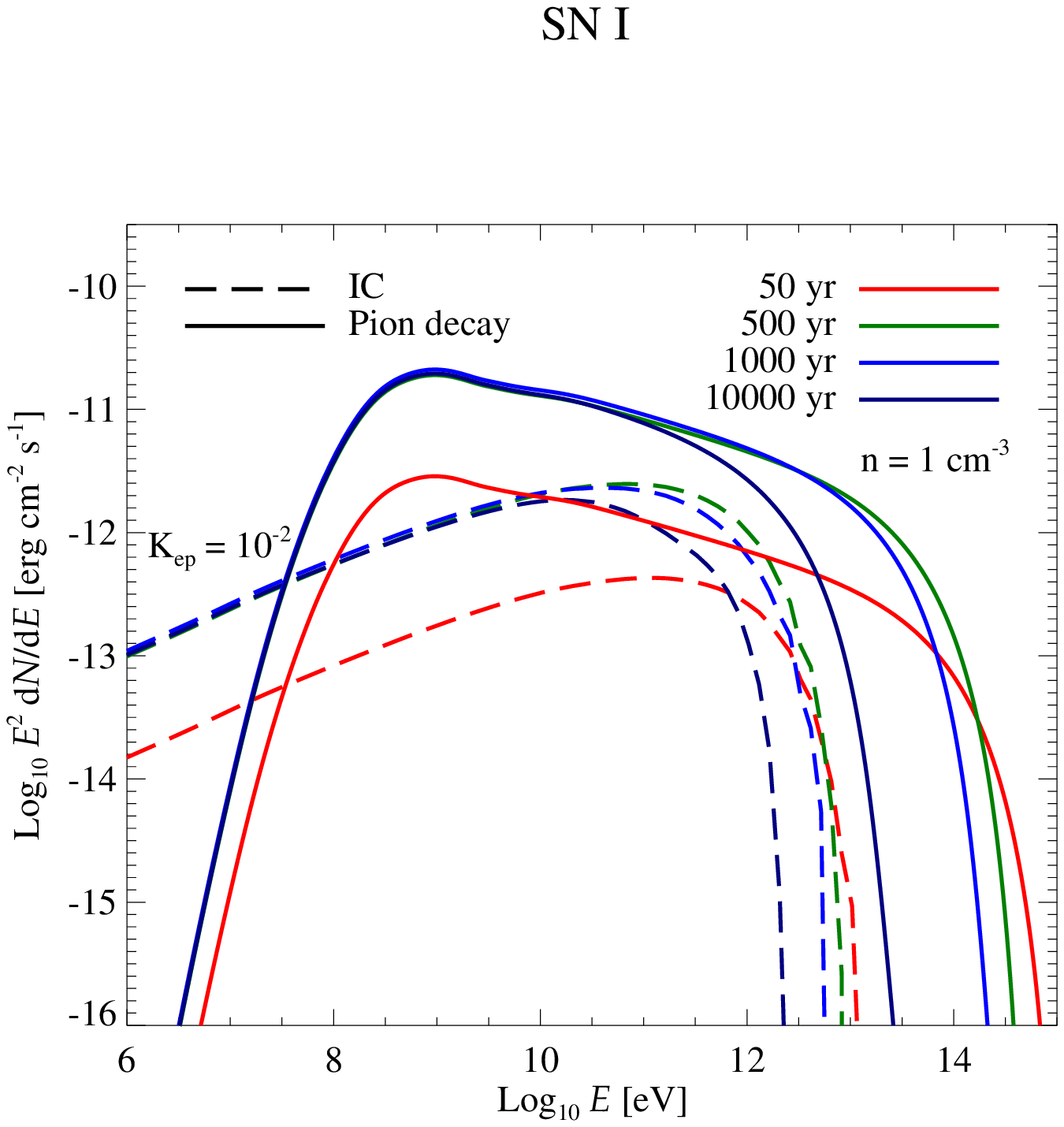}
\endminipage\hfill
\minipage{0.48\textwidth}
\includegraphics[scale=0.56]{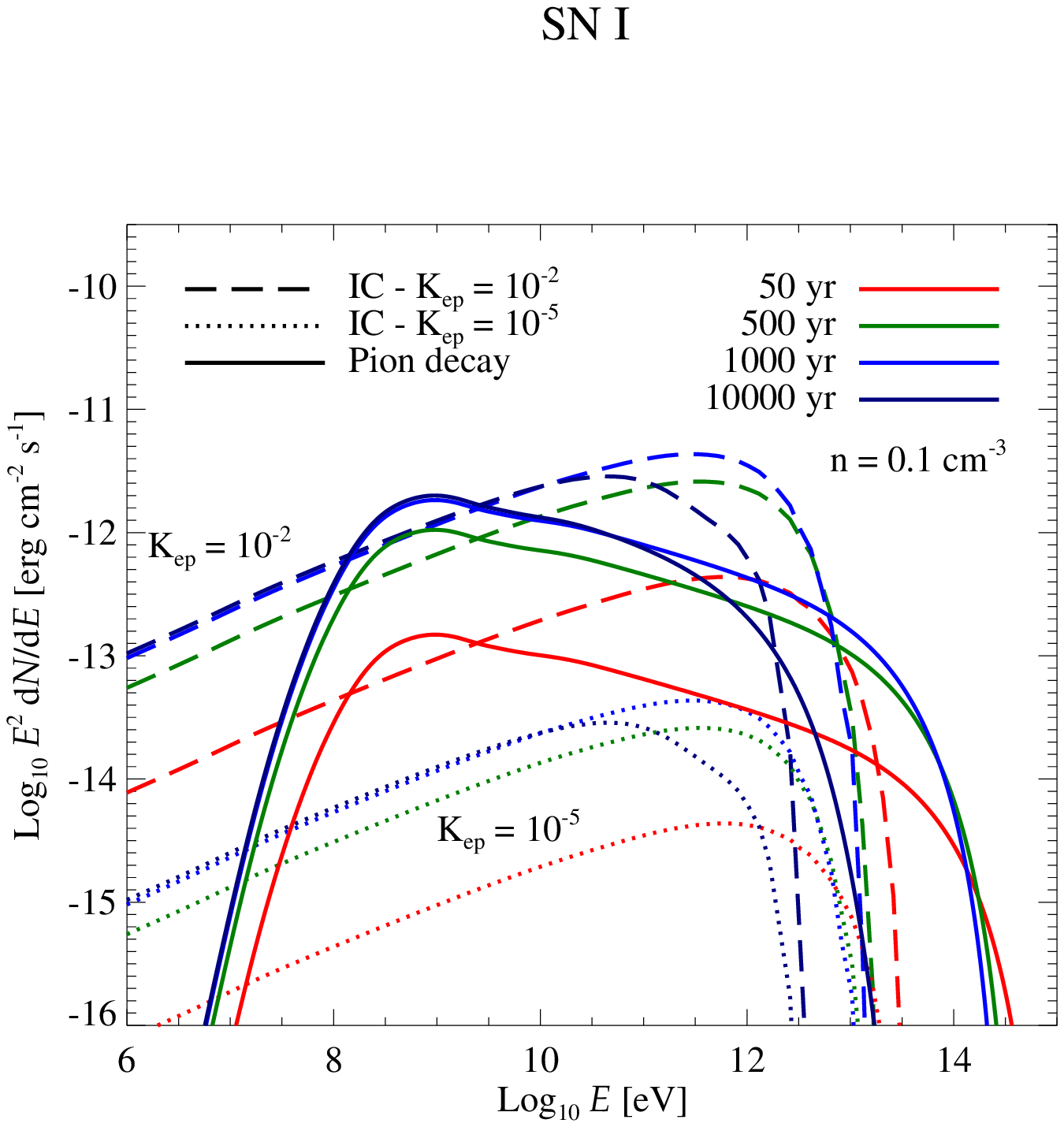}
\endminipage
\caption{Impact of the interstellar gas density and $K_{\rm ep}$ on the gamma-ray spectra of type Ia SNRs.}
\label{fig:SNI_gas_kep}
\end{center}
\end{figure*}

\begin{figure}
\begin{center}
\includegraphics[scale=0.56]{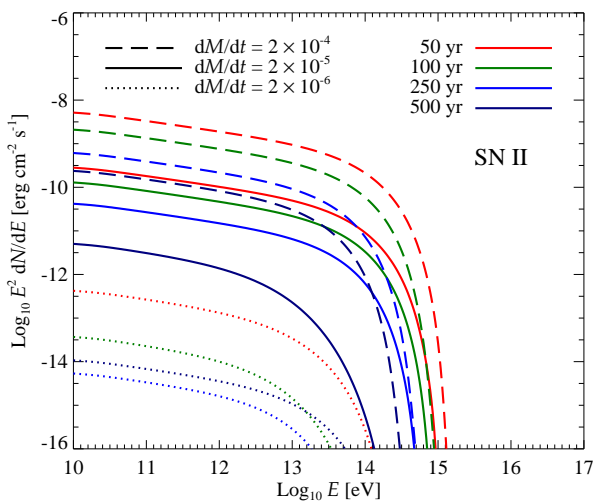}
\caption{Impact of $\dot{M}$ on the evolution of the spectra of type II SNRs.}
\label{fig:SNII_zoom_Mdot}
\end{center}
\end{figure}

\begin{figure*}
\begin{center}
\minipage{0.46\textwidth}
 \includegraphics[scale=0.56]{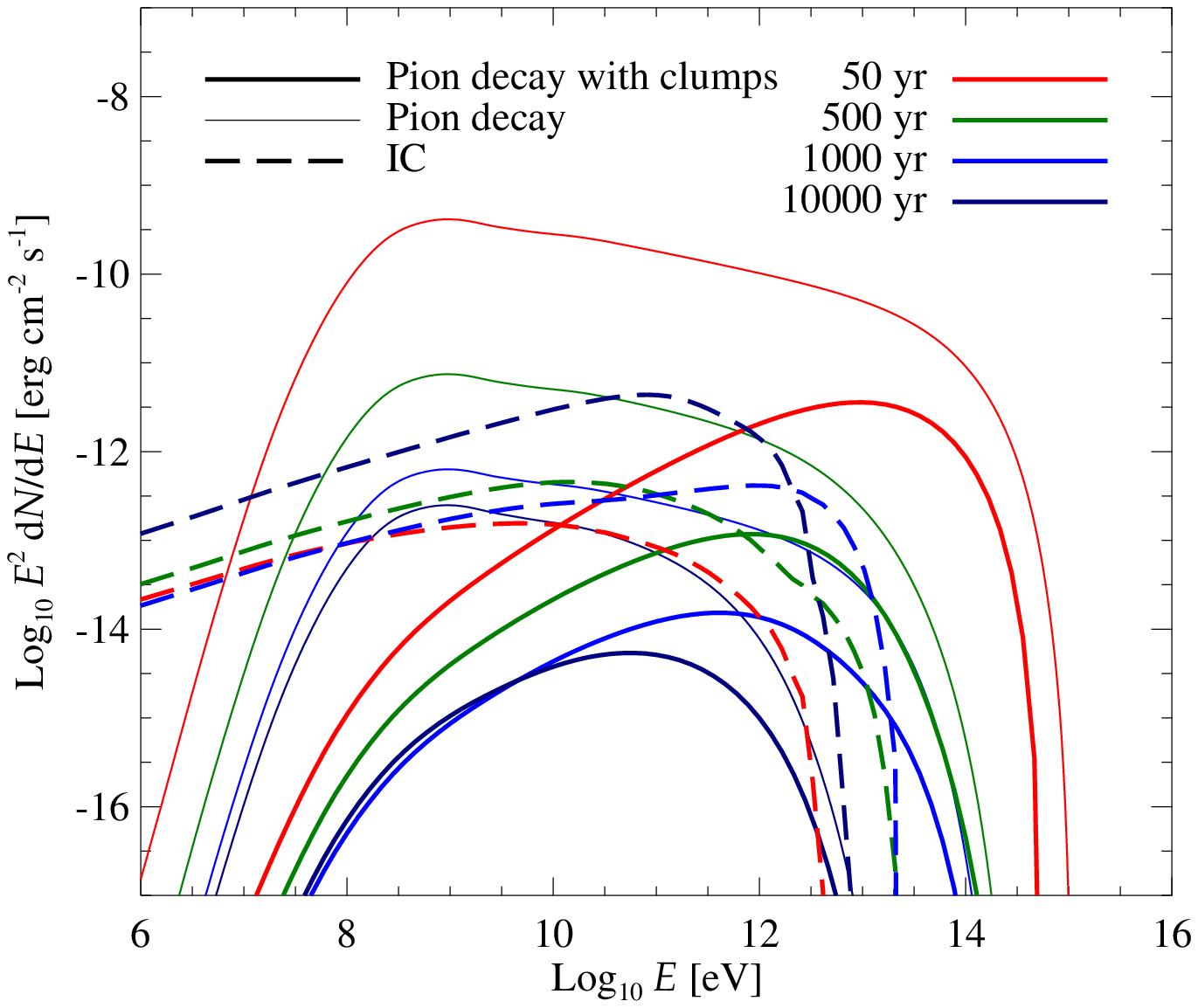}
\endminipage\hfill
\minipage{0.46\textwidth}
\includegraphics[scale=0.56]{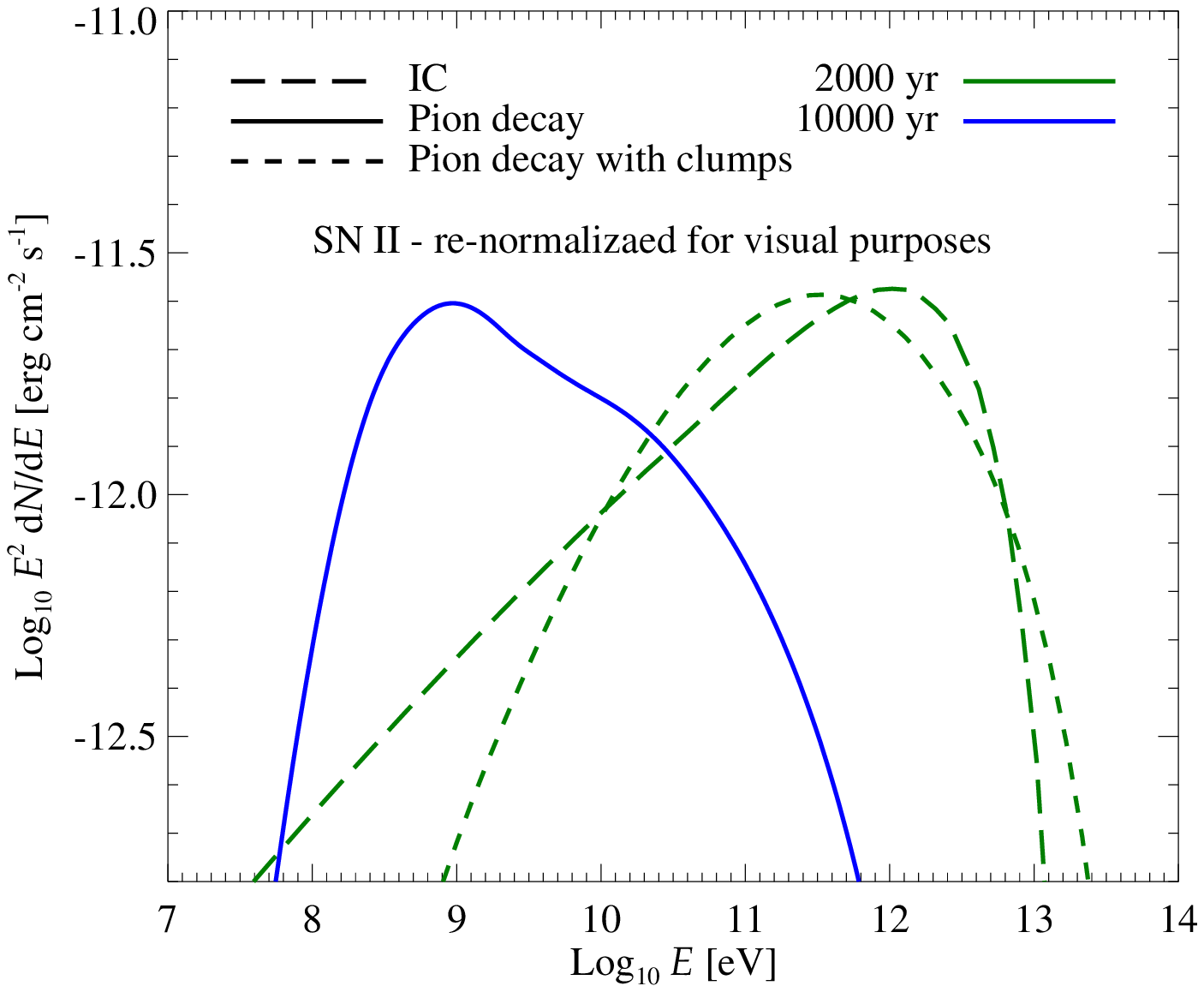}
\endminipage
\caption{Impact of the presence of clumps on the evolution of the spectra of type II SNRs on the left, and representation of the observed trend of young-IC-dominated and old-pion-decay-dominated SNRs on the right (see main text for details).}
\label{fig:SNII_zoom_trend}
\end{center}
\end{figure*}

\begin{figure*}
\begin{center}
\minipage{0.46\textwidth}
\includegraphics[scale=0.56]{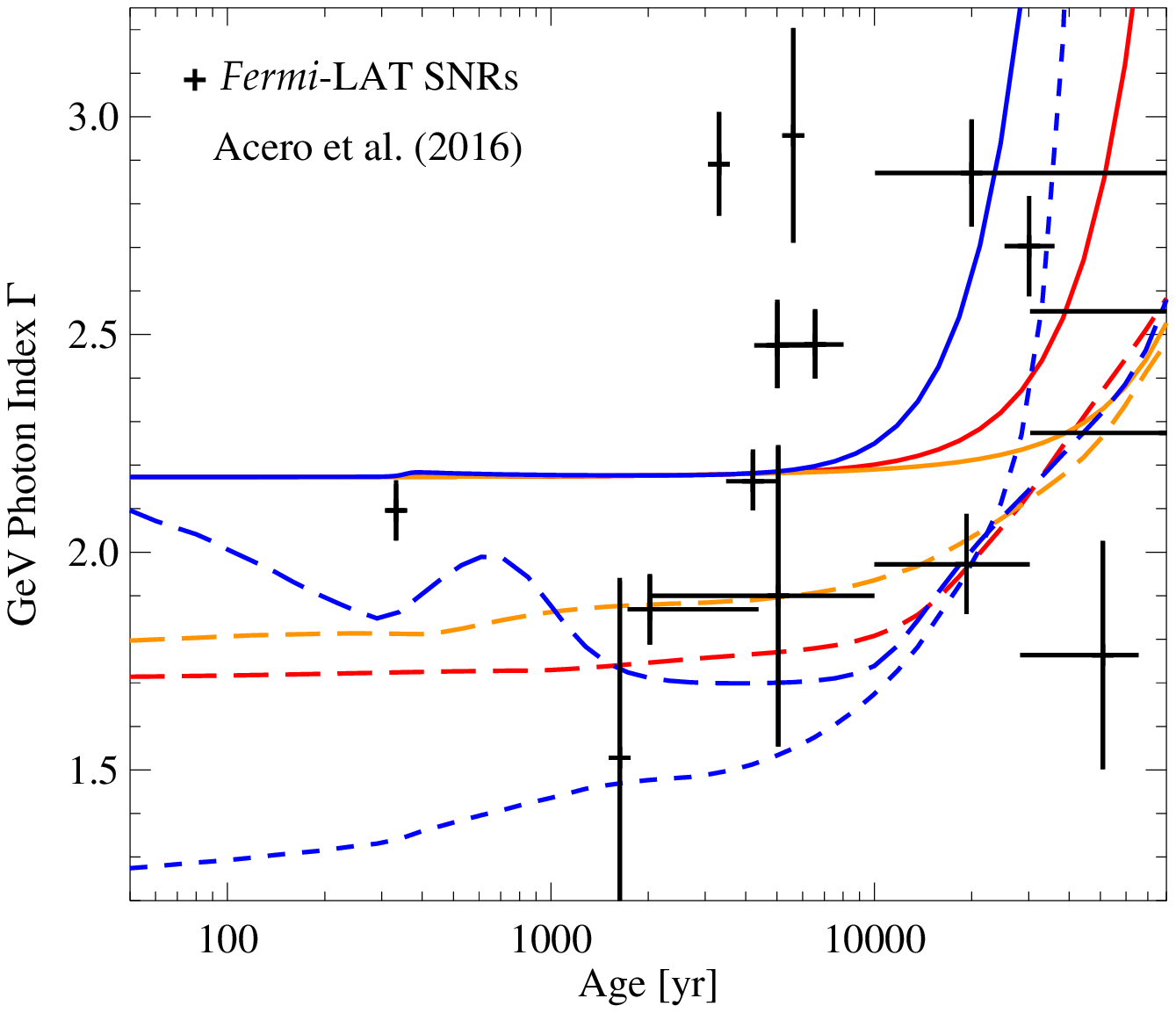}
\endminipage\hfill
\minipage{0.46\textwidth}
\includegraphics[scale=0.56]{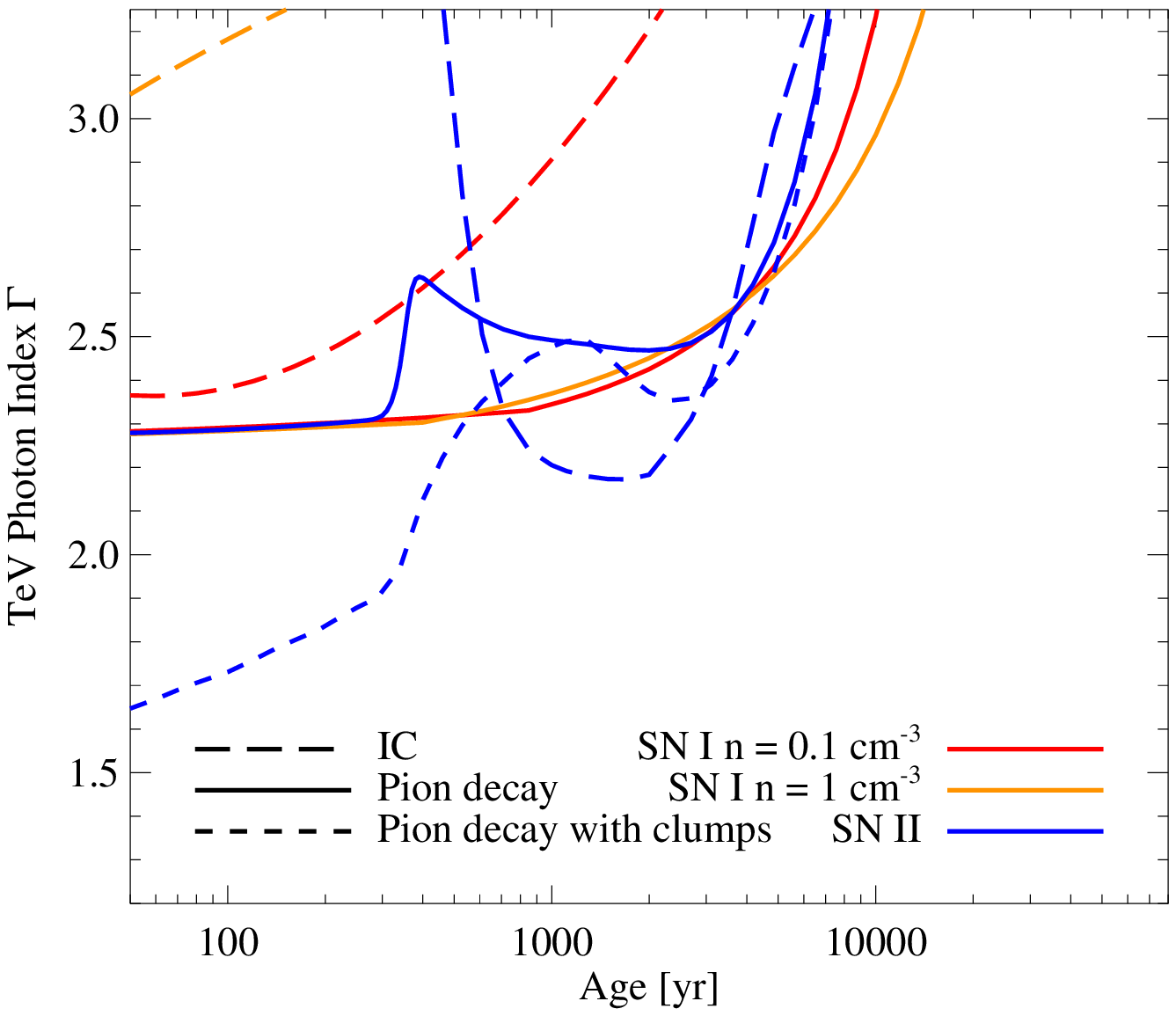}
\endminipage
\caption{Evolution of the $1-100$~GeV (left) and $0.5-5$~TeV (right) photon spectral index of IC and pion-decay emissions of type Ia and II SNRs. The legend for the curves shown in the right panel applies also to the left panel.}
\label{fig:GeVTeVsp}
\end{center}
\end{figure*}

We discuss here our results on the time evolution of the broadband emission from type I and type II remnants (assuming they are located at 2~kpc distance). Once we have computed the evolution of the electron and proton spectra within the framework described in the previous paragraph, we compute the corresponding synchrotron, inverse-Compton (IC) and pion-decay spectra. 

The synchrotron and IC spectra are computed following \cite{1970RvMP...42..237B}. The magnetic field we consider for the computation of the synchrotron spectrum is the average magnetic field within the remnant $B_{\rm inside}$. The scattered photon field for the IC computation is the inter-stellar radiation field including the cosmic microwave background (CMB), and the infrared, stellar and UV light, for which we use the analytical approximation of \cite{2010A&A...524A..51D}, based on the model described in \cite{porter}; in particular, we use the model M1 of their table~2. We remark that the CMB and the infrared components  dominate the IC emission.
The computation of the pion-decay gamma-ray spectrum is done using the \cite{2014PhRvD..90l3014K} formulation with the {\tt NAIMA} public code \citep{naima}. {In this case, at each timestep, the target density for the hadronic collisions is taken to be the swept-up mass as if it were enclosed in a shell at $0.9-1$~$R_s$}.

In Figure~\ref{fig:benchmarkCases}, we consider two benchmark cases for the evolution of a type Ia and II SNR shown in the right and left panel, respectively (we refer again to table \ref{tab:example_table} for the list of the main parameters). In particular, for our benchmark cases, we assume $n=0.1$~cm$^{-3}$ as interstellar medium density for the type Ia SNR, and we adopt an electron-to-proton fraction of $K_{\rm ep} = 0.01$. The broadband emission is shown for some relevant reference ages: $50$~yr (when the acceleration is expected to be efficient all the way up to PeV energies), $500$~yr (comparable to the age of the well-known remnants {\tt Cas A} and {\tt Tycho}), $10^3$~yr (comparable with the age of {\tt RX J1713.7-3946}), and $10^4$~yr (a late time when the remnant is at the end of the Sedov-Taylor phase).

The different spectral shapes of the synchrotron, IC and pion-decay emission are clearly seen in the figure. The well-known characteristic feature below the $\pi^0$ mass scale associated to hadronic emission (the so called {\it pion bump}) mainly depends on the ratio between the number density of electrons and protons ($K_{\rm ep}$), and the gas density of the surrounding medium. For our benchmark choice ($K_{\rm ep} = 0.01$, $n = 0.1\, {\rm cm^{-3}}$), the pion bump  is barely visible on top of the IC emission for the type Ia SNR, likely not enough to be identified observationally. However, for the type II SNR, the pion-decay  emission largely dominates over the IC at early stages, become comparable, but with a rather different spectrum, around $10^3$~yr, and eventually subdominant at the latest stages. 

The other key feature of interest in these spectra is the evolution of the high-energy cutoff. Consistently to what we showed in the previous sections, a typical SNR -- for the default values of the parameters considered here -- is a bright gamma-ray emitter all the way up to energies consistent with a CR population extending up to PeV (i.e., a PeVatron) during the early stage of its evolution, until $\sim100$~yr. We also note that some minor features can be appreciated in the synchrotron and IC spectra of the type II SNR in the right panel of Figure~\ref{fig:benchmarkCases} - these can be traced back to the shape of our evolving electron spectrum.

In Figure~\ref{fig:SNI_gas_kep}, we show how the pion bump in the type Ia SNR is highly enhanced, and, {\it vice versa}, heavily suppressed, for values of the interstellar gas density as large as $1 \,{\rm cm^{-3}}$, and values of $K_{\rm ep}$ as low as $10^{-5}$, respectively.

The presence of the PeVatron phase, and its duration, is obviously liked to the maximal energies reached by the protons. In our computation, $E_{\rm max}$ mainly depends on:
\begin{itemize}
\item the effectiveness of the magnetic field amplification ($\xi_B$) upstream of the shock, which still remains an open issue \cite{bell2013,drury2012,giacalone2007}; and 
\item the value of the gas density in the interstellar medium around the remnant, since we are assuming that $3.5$\% of the shock ram pressure $P_{\rm ram}= \rho(r) u_{\rm w}(r)$ in converted into magnetic energy.
\end{itemize}
In Figure~\ref{fig:SNII_zoom_Mdot}, we investigate the latest point, hence assuming that the magnetic field amplification mechanism is effectively operating, and show the time evolution of the hadronic gamma-ray spectrum for a type II SNR for different values of $\dot{M}$. We find that the SNR is not a PeVatron, even at very early times, for values of $\dot{M}$ lower than $\sim10^{-5} M_{\odot}/$yr (being $2\times10^{-5} M_{\odot}/$yr our benchmark value), while for higher values it safely attains the PeVatron phase up to $\sim100$~yr { (see \citealp{smith} for a comprehensive review on the mass loss rates in different types of progenitor stars)}.

Let us now turn our attention to the role of clumps. In the left panel of Figure~\ref{fig:SNII_zoom_trend}, we show the effect of the evolution in a clumpy medium, as described in Section~4.1, for the benchmark case of our type II SNR. The key feature in this case is that the resultant pion-decay gamma-ray spectra are much harder than the clumpy-free ones, mimicking the spectra typically obtainable with IC emission, and even harder for our parameter choices. We stress this point again in the right panel of Figure~\ref{fig:SNII_zoom_trend}, where we try to visualize the {trend suggested by current observations} of young IC-dominated and old pion-decay dominated SNRs  -- see e.g. fig. 6 in \cite{funk} -- for the case of our type II SNR. In this figure, we re-normalized arbitrarily the different spectra at different ages to roughly peak at the same value for visual purposes. In particular, we show the IC and pion-decay emission at $2000$ and $10^{4}$~yr as representative of a typical young and middle-age SNR, respectively. We then over-plot the pion-decay emission for a SNR evolving in a clumpy medium and see that this can well mimic the classical shape of an IC spectrum. 

As a final discussion point, we compute the photon spectral indexes for our benchmark cases for the IC and pion-decay emission in the $1-100$~GeV and $0.5-5$~TeV energy ranges. We show the resulting trends in Figure~\ref{fig:GeVTeVsp}. We remind the reader that our injected electron and proton spectral index is $\alpha = 2.3$ (see Section~4), changing this to, e.g., $\alpha = 2.1$ does not change much the trends in Figure~\ref{fig:GeVTeVsp} but for an overall shift downward of about 0.1 and 0.2 in photon index for IC and pion-decay emission, respectively. The two panels for the GeV and TeV photon spectral index allow the reader to visualize in a more complete way the spectral evolution at all times. The left panel of Figure~\ref{fig:GeVTeVsp} shows the evolution with time of the GeV photon index for our benchmark cases. We notice a regular behavior, consistent with what we have found so far. In particular, the features clearly visible in the evolution of the type II SNRs can be easily traced back to the discussion in Sections~2 and 3, and shown in Figure~\ref{fig:EmaxIa}. The same is true for the left panel where the TeV photon index behaviors are shown. These are, in general, less regular and not as obviously interpreted, but we can do so keeping in mind the IC and pion-decay spectra at these energies which are, for most of the considered time steps, on the edge of the emissions cutoff. This can indeed be seen in both panels: a rapidly growing photon spectral index means that the energies at which we are fitting for it are falling beyond the emissions cutoff. The behaviors for the type II SNRs are particularly interesting also at TeV energies, where we can clearly appreciate the importance of the IC component around $10^{3}$~yr. The effect of the evolution in a clumpy medium discussed above is clear here both at GeV and TeV energies: the photon spectral indexes of the pion-decay component can be similar, or even harder, than the ones of the IC component for most of the considered time steps.

In the left panel of Figure~\ref{fig:GeVTeVsp}, we show the data points corresponding to the SNRs in the First \emph{Fermi}-LAT Supernova Remnant Catalog (\citealp{2016ApJS..224....8A}; see their figure~16) but excluding those classified as interacting with molecular clouds as our current set-up does not model such phase. We show these only for completeness as not much can be deduced from such a comparison at the current observational stage. For the same reason, we do not include TeV data points in the right panel as most of the TeV SNRs are interacting with molecular clouds and/or have larger uncertainties in the spectral indexes. However, future more complete and accurate observational samples both from \emph{Fermi}-LAT, and hopefully its successors (see, e.g., \citealp{2017ExA...tmp...24D}), and {from CTA \citep{CTA, pierre2} will make such comparisons a powerful tool.} 

\section{Summary}

In this paper we studied the time evolution of the (hadronic and leptonic) cosmic-ray spectrum inside a typical type Ia and type II supernova remnant, and computed the associated evolution of the non-thermal emission.

To this aim, we developed a complete numerical framework mainly inspired by phenomenological considerations. The key ingredient is a significant field amplification. Motivated by the available X-ray observations of young SNRs, we modeled this process by assuming that a small, constant portion ($\simeq 3.5$\%) of the energy flux entering the remnant during the expansion of the shock wave is effectively converted into magnetic energy.

We followed the evolution of two benchmark cases, representative of a typical SNIa and SNII, and parameterized the problem in order to allow scans over the relevant parameter space. 

We focused on several key aspects: The relevance and duration of the PeVatron phase, the role of the clumpiness of the ambient medium, and the relevance of the characteristic {\it pion bump} feature in the gamma-ray spectrum, a widely used indicator of the hadronic origin of the gamma-ray emission.

We found that a SNII can sustain the PeVatron phase for several decades, during the early phase of its evolution, due to the combined effects of the large shock velocity and the high density of the ambient medium. The shock propagates in the thick progenitor wind and the magnetic field amplification, linked to the incoming flux of particles, is particularly effective. We discussed how the duration of the PeVatron phase depends on the parameters involved in the problem. 

We then computed the gamma-ray spectra at different times, and pointed out how the {\it pion bump} feature is affected by the largely unknown ratio between the number of accelerated electrons and protons, and by the properties of the ambient medium.

We then turned our attention to the role of clumps, and outlined their crucial role in shaping the gamma-ray spectrum. In particular, as a consequence of the easier penetration of high-energy CRs inside the densest clumps, we found a significant hardening of the hadronic emission from SNRs exploding in a clumpy medium, and explicitly showed the subsequent degeneracy between hadronic and leptonic spectra especially in middle-aged remnants.

Our numerical framework appears suitable for systematic scans of the parameter space and population studies, and will be useful in the prospect of a more complete collection of multi-wavelength data, in particular in the TeV domain, that will be provided in the next years by CTA. As a preliminary step, we provided a comparison between the predicted time evolution of the GeV gamma-ray spectral index and the current collection of available data, and presented the expected time evolution of the TeV slope.

\section*{Acknowledgements}

We thank D. Caprioli for useful comments.

SG acknowledges support from the Observatory of Paris (Action F\'ed\'eratrice CTA) and from the Programme National Hautes Energies (PNHE) funded by CNRS/INSU- IN2P3, CEA and CNES, France.

This work was supported by the Netherlands Organization for Scientific Research (NWO) through a Veni grant (FZ).






\bsp	
\label{lastpage}
\end{document}